\documentstyle[aps,epsfig,prl,twocolumn]{revtex}
\draft
\begin{document}
\twocolumn[\hsize\textwidth\columnwidth\hsize\csname
@twocolumnfalse\endcsname

\title{Evidence for an unconventional magnetic instability \\
 in the spin-tetrahedra system $\rm Cu_2Te_2O_5Br_2$}

\author{P. Lemmens$^1$, K.-Y. Choi$^1$, E.E. Kaul$^2$, Ch. Geibel$^2$, K.
Becker$^3$,\\
 W. Brenig$^4$, R. Valenti$^5$, C. Gros$^5$, M. Johnsson$^6$, P. Millet$^7$, F. Mila$^8$}
\address{$^1$2. Phys. Inst., RWTH Aachen, 56056 Aachen, Germany,\\
 $^2$MPI-CPfS, 01187 Dresden, Germany,\\
 $^3$Inst. f\"{u}r Theor. Physik, TU-Dresden, 01069 Dresden,
 Germany,\\
 $^4$Inst. f\"{u}r Theor. Physik, TU-Braunschweig, 38106 Braunschweig,
 Germany, \\
 $^5$Inst. f\"{u}r Theor. Physik, Univ. des Saarlands, 66041 Saarbr\"{u}cken,
 Germany, \\
 $^6$Dep. of Inorg. Chem., Stockholm Univ., 10601 Stockholm,
 Sweden, \\
 $^7$CEMES/CNRS, 31062 Toulouse, France, \\
 $^8$Inst. de Physique Theor., Univ. Lausanne, 1051 Lausanne, Switzerland}

\date{June 20th, 2001}

\maketitle

\begin{abstract}
Thermodynamic experiments as well as Raman scattering have been used to
study the magnetic instabilities in the spin-tetrahedra systems $\rm
Cu_2Te_2O_5X_2$, X=Cl and Br. While the phase transition observed in the
Cl system at $\rm T_o$=18.2~K is consistent with 3D AF ordering, the phase
transition at $\rm T_o$=11.3~K in the Br system has several unusual
features. We propose an explanation in terms of weakly coupled tetrahedra
with a singlet--triplet gap and low lying singlets.
\end{abstract}


]

\narrowtext

Reduced dimensionality of a quantum spin system in combination with
frustration leads in many cases to unconventional and interesting ground
states or magnetic phase diagrams. Prominent examples are the frustrated
and dimerized spin-1/2 chain, represented by the low-temperature phase of
$\rm CuGeO_3$, or the two-dimensional Shastry-Sutherland lattice with
orthogonally arranged spin dimers and a frustrating inter-dimer coupling,
realized in $\rm SrCu_2(BO_3)_2$. These systems show a spin liquid ground
state with a singlet--triplet gap. Frustration is evident in the latter
system as dispersionless elementary triplets and multi-particle bound
states of triplet and singlet character \cite{hase,kageyama,lemmens}.

Spin triangles and tetrahedra that are strongly coupled into Kagom\'e or
pyrochlore structures are at the origin of another important class of
frustrated spin systems \cite{hagemann,ramirez}. Although the consequences
of the classical ground-state degeneracy for the quantum case have not
been fully elucidated theoretically, there are good reasons to believe
that such models do not possess magnetic long-range order (LRO) but
low-lying singlets \cite{chandra,mila}. Scenarios leading to the
development of LRO within this non-magnetic manifold have been put forward
\cite{kotov,ueda}.

The limit of weakly-coupled tetrahedra with spin S=1/2 has been
studied in some detail in 3D and 1D, and the physics is expected
to be very interesting\cite{harris,canals,tsu,brenig}. However,
they have not been investigated experimentally so far due to the
lack of appropriate materials. The recently found spin system $\rm
Cu_2Te_2O_5X_2$, with X=Cl, Br, contains tetrahedral clusters of
$\rm Cu^{2+}$ with S~=~1/2 in a distorted square planar $\rm
CuO_3X$-coordination \cite{johnsson}. These tetrahedra align to
tubes or chains along the [001] direction, as they are separated
along [100] and [010] direction by different Te-O coordinations
(see Fig. 1a). Substituting Br for Cl in this system widens up the
unit cell and increases its volume from 367 to 391$\AA^3$, by 7\%,
while bond angles or anisotropies do not change essentially.
Therefore, this system allows in a unique way to study the
interplay of frustration and coupling in a tetrahedra quantum spin
system.

Preliminary measurements of the magnetic susceptibility $\chi$(T)
of both compounds showed a maximum at $\rm T_{\chi_{max}}$~=~23~K
and 30~K for X=Cl, Br and a strong reduction at low temperatures,
typical for a spin gap system. Assuming that the compounds consist
of weakly-coupled units of 4 spins with couplings $\rm J_1$ and
$\rm J_2$ (see Fig. 1b), the best fit of the susceptibility -
quite a good one actually - was obtained for $\rm J_1=\rm J_2=$
38.5 K and 43 K for X=Cl and Br respectively. The effect of
inter-tetrahedra coupling was not included in this fit however
\cite{johnsson}.

We present here an investigation of both compounds, using detailed
susceptibility, specific heat and Raman measurements. These
measurements indicate an onset of antiferromagnetic order in the
chloride compound at 18.2~K, pointing to a significant
inter-tetrahedra coupling. In contrast, the bromide shows a very
unusual phase transition at 11.3~K, at a temperature where a large
part of the magnetic excitations have already been frozen out. A
possible scenario for this transition shall be discussed.

The preparation of the samples and single crystals used for our
measurements is described elsewhere \cite{johnsson}. Specific heat
and susceptibility measurements were performed on powder samples
using a Quantum Design measurement property system and a SQUID
magnetometer, respectively. The Raman scattering experiments were
performed with a $\lambda$~=~514~nm laser line and a power level
of 0.05-2~mW focused on a spot of 0.05-0.1~mm diameter. The
whisker-like single crystals with typical dimensions $\rm 0.1\cdot
0.1\cdot 1~mm^3$ and the measurement geometry did allow
experiments in (cc), (yc) and (yy) light scattering polarizations,
with c parallel to the crystallographic [001] direction and y
given by a linear combination of the [100] and [010] direction.
Based on the tetragonal $\rm P\overline{4}$ space group and the
missing center of inversion these polarizations correspond to the
\textbf{A} symmetry and combinations of \textbf{B} and \textbf{E}
symmetry, respectively.

Results of our detailed investigation of the magnetic
susceptibility $\chi$(T) of both compounds are shown in Fig. 2.
Whereas the high temperature part of $\chi$(T) at and above $\rm
T_{\chi_{max}}$ is very similar in both compounds, pronounced
differences are observed at low temperatures. In the chloride a
clear kink is evident in $\chi$(T) at $\rm T_N$ = 18.2~K. This
anomaly corresponds to a pronounced step-like increase of the
slope $\partial\chi$(T)/$\partial$T with decreasing temperature,
as shown in the inset of Fig.2. In other words, the susceptibility
in the ordered state is smaller than it would be in the disordered
state. Applying a magnetic field leads to a reduction of the size
of this anomaly and thus to an increase of the susceptibility for
$\rm T<T_N$.

The specific heat $\rm C_p$(T) of $\rm Cu_2Te_2O_5Cl_2$ (see Fig.
3, upper curves with upper temperature scale) shows a mean field
like transition with a sizeable anomaly at $\rm T_N$. A magnetic
field of 13.5~T leaves this anomaly almost unchanged, only a
slight decrease of $\rm T_N$ from 18.2~K to 18.0~K is discernible.
All these features: the reduction of the susceptibility for $\rm
T<T_N$, the mean field type of transition, the reduction of the
anomaly in $\chi$(T) in an applied field whereas no reduction is
observed in $\rm C_p$(T) and the very weak decrease of $\rm T_N$
with increasing B, point to a 3D-antiferromagnetic ordering in a
system with only weak spin-anisotropies. Since this transition
occurs below the well-defined maximum in $\chi$(T), with $\rm
T_N$/$\rm T_{\chi_{max}}$ = 0.78, these results indicate that $\rm
Cu_2Te_2O_5Cl_2$ is a low-dimensional spin systems with a
significant inter-tetrahedra (or inter-chain) coupling.

In contrast, the low temperature behavior of $\rm Cu_2Te_2O_5Br_2$ is
quite different and rather unusual. For temperatures below $\rm
T_{\chi_{max}}$, $\chi$(T) decreases considerably, by more than 50~\%.
This indicates the freezing out of a large part of the magnetic triplet
excitations as expected in a system with a spin gap. Whereas at first
sight no anomaly can be observed in $\chi$(T) at low fields (0.1~T), the
derivative $\rm
\partial\chi (T)$/$\rm \partial T$ reveals a small but well
discernible step at $\rm T_o$ = 11.5~K. This step however has the
opposite sign compared to $\rm Cu_2Te_2O_5Cl_2$, i.e. the slope
for $\rm T<T_o$ is smaller than for $\rm T>T_o$. This means, that
the susceptibility of $\rm Cu_2Te_2O_5Br_2$ and thus the
magnetization in the ordered state is larger than it would be in
the disordered state. This is opposite to the expected result for
antiferromagnetic ordering. For $\rm T<T_o$, the magnetization is
strictly proportional to the applied field between B~=~-1~T and
1~T and shows no remanence. Thus ferromagnetic ordering or canted
antiferromagnetism can be excluded for $\rm Cu_2Te_2O_5Br_2$. The
anomaly in $\chi$(T) increases significantly in a magnetic field
larger than 1~T. For B~=~5~T, it is quite evident that $\chi$(T)
is higher in the ordered state than it would be in the disordered
state.

The specific heat of  $\rm Cu_2Te_2O_5Br_2$ at B~=~0 (see Fig. 3,
lower curves with lower temperature scale) shows a small but
well-defined mean field like anomaly at $\rm T=T_o$. This proves
that this transition also occurs in the absence of an external
field. Applying a magnetic field leads to a very strong increase
of the size of the anomaly, by more than a factor of 3 at
B~=~13~T, and to a pronounced shift of $\rm T_o$ to higher
temperatures, from $\rm T_o$(0)~=~11.4~K to $\rm
T_o$(B=13T)~=~12.4~K. In thermodynamic terms, this shift
corresponds to the larger magnetization of the ordered state
compared to the disordered state. Fitting a power law to the field
dependence of $\rm T_o$ as determined from $\rm C_p$(T), $\rm
T_o(B)=T_o(0)$+$\rm a \cdot B^{n}$, we obtain $\rm n=1.41\pm
0.05$. Using the low temperature parts (T$\ll$$T_o$, $\rm T_N$)
and the high temperature parts (T$>$T$\rm _o$, $\rm T_N$) of the
specific heat results, we made a rough estimate of the magnetic
specific heat and of the magnetic entropy $\rm S_m(T)$. For $\rm
Cu_2Te_2O_5Br_2$, the magnetic entropy at $\rm T_o$ is only a
quite small portion of the total spin entropy expected at high
temperatures $\rm S_m(T_o) \simeq 1.8~J/Kmol = 0.16 \cdot
Rln2/spin$, whereas for $\rm Cu_2Te_2O_5Cl_2$, the entropy at $\rm
T_N$ is much larger, $\rm S_m(T_N) \simeq 4.1~J/Kmol = 0.36 \cdot
Rln2/spin$. This is related to the difference in the ratio $\rm
T_o/J$ and $\rm T_N$/J and indicates that in the bromide, a large
part of the magnetic degrees of freedom are freezing out at higher
temperatures.

Light scattering studies have been performed on both systems as
function of temperature. The optic (q$\approx$0) phonon
frequencies of $\rm Cu_2Te_2O_5Br_2$ are generally smaller
compared with $\rm Cu_2Te_2O_5Cl_2$. This is consistent with the
larger unit cell volume of the bromide. Details of the phonon
spectrum will be given elsewhere. In the low energy range
comparable to the spin gap of the systems we observe two signals
in $\rm Cu_2Te_2O_5Br_2$ that do not fit to phonon scattering. The
excitation spectrum of $\rm Cu_2Te_2O_5Cl_2$ is less spectacular.
The corresponding energy regime displays a much weaker temperature
dependence and a smaller intensity of scattering. It will not be
further discussed here.

In $\rm Cu_2Te_2O_5Br_2$ the high-temperature spectra are
dominated by a pyramidal-shaped scattering continuum centered at
61~$\rm cm ^{-1}$=88~K, corresponding to 2$\Delta$~=~86~K
previously determined from the magnetic susceptibility
\cite{johnsson}. The continuum is attributed to a two-magnon-like
scattering process \cite{brenig}. Its total linewidth and the low
energy onset at 40~$\rm cm ^{-1}$ both point to an appreciable
inter-tetrahedra coupling. For reduced temperatures, $\rm T<9~K$,
a second maximum with smaller linewidth develops. It shows a soft
mode-like behavior and reaches its maximum energy of 18~$\rm cm
^{-1}\equiv 0.6\Delta$ at lowest temperatures. This intensity is
undoubtedly related to the instability observed in our
thermodynamic experiments. In the same temperature regime an
additional shoulder develops on the high frequency side of the
scattering continuum. This leads to a small shift of this signal
from 61 to 63~$\rm cm ^{-1}$.

These two scattering signals are only observed in (cc)
polarization with both electric field vectors parallel to the
crystallographic c-axis and the chains of tetrahedra. This
symmetry selection rule and the temperature dependence of the low
energy mode are similar to the properties of a singlet bound state
observed in the dimerized phase ($\rm T<T_{SP}~=~14.5~K$) of $\rm
CuGeO_3$. In this spin chain system frustration leads to a binding
effect of two elementary triplets to a well-defined mode at $\rm
1.78\Delta$ \cite{muthukumar}. The "in chain" selection rule that
is observed in both systems in combination with the final
linewidth of the continuum are the result of an appreciable
quasi-one-dimensional inter-tetrahedra coupling.

Let us start the discussion by noting that the main features of the Br
system can be understood in terms of weakly coupled tetrahedra. The
spectrum of a tetrahedron is depicted in Fig. 1c. If $\rm J_1=\rm J_2$
above $\rm T_o$, as suggested by the fit of the susceptibility at high
temperature, then the ground-state is a two-fold degenerate singlet, and
the first excitation is a three-fold degenerate triplet located at
$\Delta=\rm J_1$ above the GS. These triplets would then lead to Raman
scattering at $2\Delta=86~K$, in agreement with our data.

The origin of the phase transition at $\rm T_o$ could be a small
distortion of the tetrahedra that leads to different values for $\rm J_1$
and $\rm J_2$, hence to a lifting of the degeneracy between the two GS
singlets. This is similar to the spin-Peierls scenario of Yamashita and
Ueda\cite{ueda}. In this picture, the Raman scattering observed below 9~K
would involve transitions between the singlet GS and the first excited
singlet. Note that the entropy jump of such a transition is expected to be
a small fraction of $\rm R\ln 2/spin$, in agreement with the specific heat
data.

Let us now discuss all the other issues raised by the present
results. First of all, the Cl and Br compounds turn out to be very
different. As a first step toward understanding this difference,
we have performed ab-initio calculations of the electronic
structure of these systems. These results will be reported
elsewhere. These calculations give access to the hopping integrals
but not to the exchange integrals. So the information is only
qualitative. The main difference between the two systems is that
the ratio of the intraplaquette hoppings $t_2/t_1$ is closer to 1
in the Br system than in the Cl one. This is consistent with our
proposal that the physics of weakly coupled tetrahedra is realized
in the Br system but not in the Cl one.

As a first step to understand the Raman continuum we have analyzed
the light-scattering cross section of a coupled array of tetrahedra,
extending however only in 1D along the c-axis \cite{brenig}. Based
on exact diagonalization and bond-operator theory it is found that
in this case the continuum is determined by the spin-zero two-triplet
excitations of an effective dimerized spin-one chain. Due to triplet
interactions the corresponding Raman continuum is strongly renormalized
with respect to the bare two-triplet spectrum. It displays no van-Hove
type edge-singularities, a width set by the inter-tetrahedral coupling,
and a smooth convex shape with a single maximum. The latter shape is
encouraging, yet still
different from the pyramid-like spectrum of fig.\ref{ramspec}. This
is consistent with our ab-initio band structure calculations indicating
that the inter-tetrahedral couplings are comparable along {\em and}
perpendicular to the tubes. A study of the Raman response of a 3D
array of coupled tetrahedra is in progress.

Finally a lot remains to be done to understand in detail what
happens at $\rm T_o$ in the Br system, in particular whether there
is a change of the local geometry, as implied by our
interpretation of Raman data, as well as the influence of the
magnetic field on the transition, which might require to take into
account the Dzyaloshinskii-Moriya interaction which is a priori
present in that system.

In summary, our results have revealed that both the Cl and Br
compounds undergo a phase transition at low temperatures, but that
these phase transitions appear to be of different nature: While
there is good evidence in favor of a 3D magnetic ordering in Cl,
the transition in the Br compound is very unusual, with the
appearance of low energy Raman scattering below the transition. If
our interpretation of the origin of this low-energy scattering is
confirmed, this makes $\rm Cu_2Te_2O_5Br_2$ the first system with
only singlet low-energy excitations that undergo a phase
transition involving these singlet low-energy degrees of freedom.
Since the Cl and Br systems are very similar a priori, their
difference at low temperature suggest that they might be located
on both sides of a quantum critical point in a system of coupled
tetrahedra.

We are grateful for helpful discussions with M. Zhitomirsky, K.
Ueda, H. Kageyama, H. Rosner, B. B\"{u}chner, and G. G\"{u}ntherodt. This
work was supported in part by the DFG through SPP1073.

\begin{figure}
\centerline{\epsfig{figure=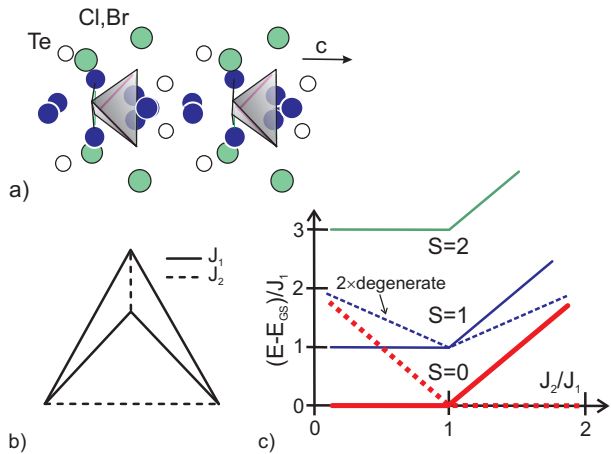,width=8cm}}
\hspace{5cm} \caption{a) Part of the crystal structure
of $\rm Cu_2Te_2O_5X_2$ with two Cu tetrahedra, O (filled), Cl or
Br (dashed) and Te (empty circles) coordinations. b) exchange
topology of a single spin tetrahedron and c) the resulting
excitation spectrum with respect to the ground state (GS) energy.}
\end{figure}


\begin{figure}
\centerline{\epsfig{figure=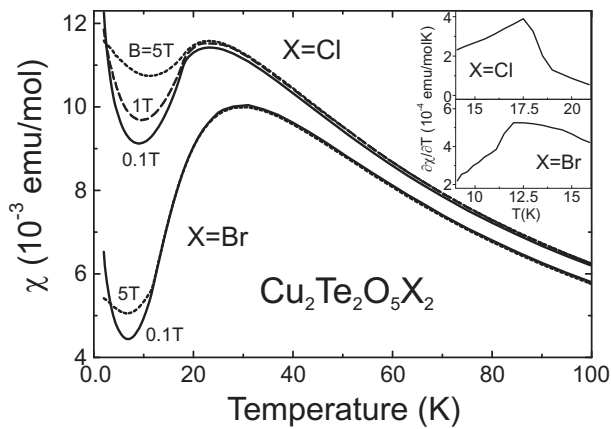,width=8cm}}
\hspace{5cm} \caption{Magnetic susceptibility
$\chi$(T) of $\rm Cu_2Te_2O_5X_2$, X=Cl and Br, with different
magnetic fields applied. The inset shows $\rm
\partial\chi (T)/ \partial T$. }
\end{figure}


\begin{figure}
\centerline{\epsfig{figure=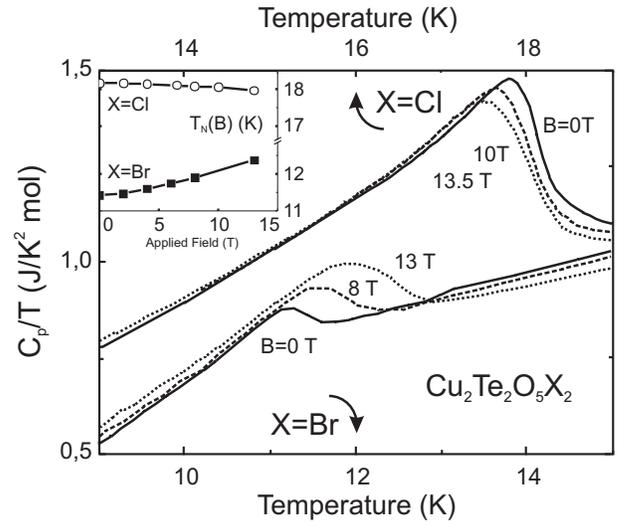,width=8cm}}
\hspace{5cm} \caption{Specific heat $\rm C_p$(T)/T of $\rm
Cu_2Te_2O_5X_2$ with X=Cl (upper temperature axis) and X=Br (lower
temperature axis). The inset gives a temperature--magnetic field
phase diagram. }
\end{figure}


\begin{figure}
\centerline{\epsfig{figure=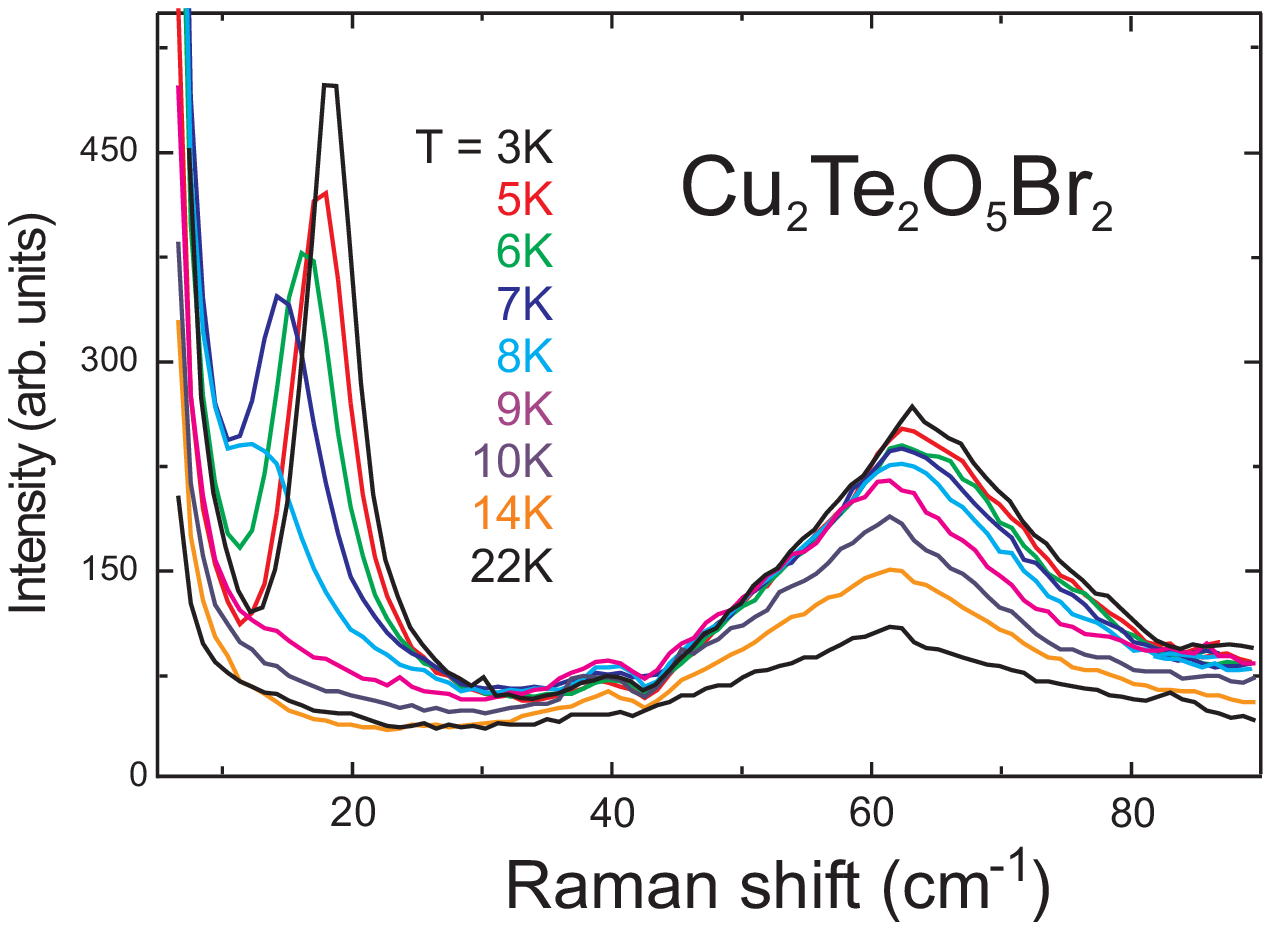,width=8cm}}
\hspace{5cm} \caption{Raman scattering spectrum of
$\rm Cu_2Te_2O_5Br_2$ in (cc) polarization. } \label{ramspec}
\end{figure}

\end{document}